\documentstyle[12pt]{article}
\begin{document}
\title{COMMENTS ON THE PAPER "ON THE UNIFICATION OF THE FUNDAMENTAL
FORCES..."}
\author{B.G. Sidharth\\ Centre for Applicable Mathematics \& Computer Sciences\\
B.M. Birla Science Centre, Hyderabad 500 063}
\date{}
\maketitle
\begin{abstract}
In this brief paper we justify observations made in El Naschie's paper
"On the Unification of the Fundamental Forces...", on the Planck scale,
fractal space time and the unification of interactions, from different
standpoints.
\end{abstract}
In a recent paper El Naschie\cite{r1} has emphasized the intimate connection
between the Planck length, the Compton wavelength and a unification of the
fundamental interactions within the framework of complex time and a fractal
Cantorian space. We will now make some observations which justify the
contentions made in the above paper, from different standpoints.\\
1. Complex Time: It has been pointed out\cite{r2,r3,r4} that Fermions can be
thought of as Kerr-Newman Black Holes, in the context of quantized space time:
There are minimum space time intervals, and when we average over these, Physics
arises. Within the minimum intervals, we encounter unphysical Zitterbewegung
effects, which also show up as a complexification of coordinates\cite{r5} - indeed
they are the double Weiner process discussed by Abbott and Wise, Nottale
and others. It may be mentioned that the transition from the Kerr metric
in General Relativity to the Kerr-Newman metric is obtained by precisely
such a complex shift of coordinates, a circumstance which has no clear
meaning in Classical Physics\cite{r6}. On the other hand it is this "Classical"
Kerr-Newman metric which describes the field of an electron including the
Quantum Mechanical anomalous gyro magnetic ratio, $g=2$. This has been
discussed in detail in references\cite{r2,r3}.\\
2. The Unification of Electromagnetism and Gravitation and the Planck
Scale: It is in the context of point 1 that we arrive at a unified picture
of electromagnetism and gravitation (Cf.ref.\cite{r4}). The point is that
at the Compton wavelength scale we have purely Quantum Mechanical effects
like Zitterbewegung, spin half and electromagnetism, while at the Planck
scale, we have a purely classical Schwarschild Black Hole. However the
Planck scale is the extreme limit of the Compton scale, where electromagnetism
and gravitation meet (Cf.ref.\cite{r4}, and \cite{r7}). This is because
for a Planck mass $m_P \sim 10^{-5}gms$ we have
\begin{equation}
\frac{Gm^2_P}{e^2} \sim 1\label{e1}
\end{equation}
whereas for an elementary particle like an electron we have the well known
equation
\begin{equation}
\frac{Gm^2}{e^2} \sim \frac{1}{\sqrt{N}} \equiv 10^{-40}\label{e2}
\end{equation}
where $N \sim 10^{80}$ is the number of elementary particles in the universe.\\
Another way of expressing this result is that the Schwarzschild radius for the
Planck mass equals its Compton wavelength. This is where Quantum Mechanics
and Classical Physics meet.\\
This point can be analysed further\cite{r7}. From equations (\ref{e1}) and
(\ref{e2}) it can be seen that we obtain the Planck mass, when the number
of particles in the universe is 1. Indeed as has been pointed out by Rosen\cite{r8},
the Planck mass can be considered to be a mini universe, in the context of
the Schrodinger equation with the gravitational interaction.\\
3. The above brings us to another interesting aspect discussed by El Naschie
in\cite{r1}. This is the fact that the Planck mass is intimately related
to the Hawking radiation, and infact from the latter consideration we can
deduce that a Planck mass "evaporates" within about $10^{-42}secs$, which also
happens to be its Compton time!\\
On the other hand, as pointed out in\cite{r5,r7}, an elementary particle like
the pion is intimately related to Hagedorn radiation which leads to a life
time of the order of the age of the universe.\\
The above two conclusions have been obtained on the basis of a background
Zero Point Field, the Langevin equation and space time cut offs leading
to a fluctuational creation of particles at the Planck scale and the Compton
scale respectively.\\
4. Resolution and the Unification of Interactions: El Naschie\cite{r1}
has referred to the fact that there is no apriori fixed length scale (the
Biedenharn conjecture). Indeed it has been argued by the author in the above
context\cite{r9,r10} that depending on our scale of resolution, we encounter
electromagnetism well outside the Compton wavelength, strong interactions at the
Compton wavelength or slightly below it and only gravitation at the Planck
scale. The differences between the various interactions are a manifestation
of the resolution.\\
4. The Universe as a Black Hole: As pointed out in\cite{r1} by El Naschie and
the author\cite{r11,r12}, the universe can indeed be considered to be a
black hole. Prima Facie this is clear from the fact that the radius of the
universe is of the order of the Schwarzchild radius of a black hole with the
same mass as the universe. Also as pointed out in\cite{r11}) the age of
the universe coincides with the time taken by a ray of light to travel from the
horizon of a black hole to its centre or vice versa.\\
5. The "Core" of the Electron: El Naschie refers to the core of the electron
$\sim 10^{-20}cms$, as indeed has been experimentally noticed by Dehmelt and
Co-workers\cite{r13}. It is interesting that this can be deduced in the
context of the electron as a Quantum Mechanical Kerr-Newman Black Hole.\\
It was shown that\cite{r2} for distances of the order of the Compton
wavelength the potential is given in its QCD form
\begin{equation}
V \approx - \frac{\beta M}{r} + 8\beta M (\frac{Mc^2}{\hbar})^2 .r\label{e3}
\end{equation}
For small values of $r$ the potential (\ref{e3}) can be written as
\begin{equation}
V \approx \frac{A}{r} e^{-\mu^2 r^2}, \quad \mu = \frac{Mc^2}{\hbar}\label{e4}
\end{equation}
It follows from (\ref{e4}) that
\begin{equation}
r \sim \frac{1}{\mu} \sim 10^{-21}cm.\label{e5}
\end{equation}
Curiously enough in (\ref{e4}), $r$ appears as a time, which is to be expected
because at the horizon of a black hole $r$ and $t$ interchange roles.\\
One could reach the same conclusion, as given in equation (\ref{e5}) from
a different angle. In the Schrodinger equation which is used in QCD, with the potential given
by (\ref{e3}), one could verify that the wave function is of the type
$f(r).e^{-\frac{\mu r}{2}}$, where the same $\mu$ appears in (\ref{e4}). Thus,
once again one has a wave packet which is negligible outside the distance
given by (\ref{e5}).\\
It may be noted that Brodsky and Drell\cite{r14} had suggested from a very
different viewpoint viz., the anomalous magnetic moment of the electron,  that
its size would be limited by
$10^{-20}cm$. The result (\ref{e5}) was experimentally confirmed by Dehmelt and co-workers
\cite{r13}.\\
Finally, it may be remarked that it is the fractal double Weiner process
referred to earlier that leads from the real space coordinate, $x$ say, to the
complex coordinate $x+\imath ct$ (Cf.ref.\cite{r5}), which is the space and
time divide: As pointed out by Hawking\cite{r15} and others, an imaginary
time would lead to a "static" Euclidean four geometry, rather than the
Minkowski world, a concept that has been criticised by Prigogine\cite{r16}. It
is in the above fractal formulation (Cf.ref.\cite{r5}), on the contrary, that
we see the emergence of the space and time divide, that is, time itself.\\
Thus in conclusion, it may be said that the recognition of a fractal quantized
underpinning of space time ties together several apparently disparate facts.

\end{document}